\documentclass[epj]{svjour}
\usepackage{graphics}
\usepackage{graphicx}
\usepackage{citesort}
\begin{document}
\title{Deuterium-deuterium nuclear cross-sections in insulator and metallic
environments}
\author{David Salzmann\inst{1} \and Michael Hass\inst{2}
\thanks{\emph{Email:} Fnhass@weizmann.ac.il}%
}                     
%
%
\institute{Marganit 2, Ness Ziona 74051, Israel \and Department of
Particle Physics, Weizmann Institute of Science, Rehovot, Israel}

\date{Received: date / Revised version: date}

\abstract{ The three-dimensional Thomas-Fermi (TF) model is used to
simulate the variation of the $d+d\rightarrow t+p$ cross-section
at low impact energies, when the target deuterium nucleus is
embedded in metallic or insulator environments. Comparison of the
 computational results  to recent experiments demonstrates that even though
 the TF model can
explain some increase in the low energy cross section for metallic
host, a full explanation of the experimental results is still lacking.
Possible reasons for the disagreement are discussed.  
} 
\maketitle
\section{Introduction}\label{sec:1}
In a series of recent experiments
\cite{rai04,rai02,cze01,yuk98,rai02a,bon03} significant
differences had been found between the low-energy cross sections
of the $d+d\rightarrow t+p$ reaction when the target nuclei are
embedded in a metallic or an insulator environment. Apparently,
the source of this effect is the conduction electron distribution
in the metallic lattice and its screening effects on the Coulomb
barrier around the incident and target nuclei \cite{ass87,ass87a}.
Hints for similar behavior were found also in the electron capture
rate of Be$^{7}$ \cite{nir07,wan06}. While in some of these
experiments temperature dependence was also claimed
\cite{rai04,rai02}, no such dependence was found in others
\cite{nir07,sev07}.

This phenomenon can be briefly introduced as follows: The
transparency, $T(E_{k})$, of a potential barrier to an incident
particle having energy $ E_{k}$ is given by \cite{fer49},
\begin{equation}\label{eq1}
\ T(E_{k})=\exp \left\{ -2\int_{r_{N}}^{b}\sqrt{\frac{2M}{\hbar ^{2}}%
\left[ E_{p}(r)-E_{k}\right] \,}dr\right\}
\end{equation}
where $M$ is the reduced mass of the two particles, $r_{N}$ - the
nuclear radius, $b$ - the classical turning point (CTP) and
$E_{p}(r)=ze\,V(r)$ is the potential energy of the incident
particle in the potential, $V(r)$, generated by the target
nucleus. $Ze$ and $ze$ are charges of the target and incident
nuclei, respectively.

If a pure Coulomb potential is substituted for the interaction potential, $%
V(r)=V_{C}(r)=Ze\,/\,r,$ Eq. (1) reproduces the Sommerfeld factor,

\begin{eqnarray}\label{eq2}
T_{C}(E_{k}) &=& \exp \left\{
-2\int_{r_{N}}^{b}\sqrt{\frac{2M}{\hbar
^{2}}\left[ \frac{Zze^{2}}{r}-E_{k}\right] \,}dr\right\} \\
  &=& \exp \left\{ -2\pi \eta
\right\} \nonumber
\end{eqnarray}
\begin{equation}\label{eq3}
\eta (E_{k})=\,\frac{Zz\,e^{2}}{\hbar v_{k}}
\end{equation}
In (\ref{eq3}) $v_{k}=\sqrt{2E_{k}/M}$ is the velocity of the
incident particle. If the target nucleus is embedded in a solid
target, the potential generated by nearby electrons, in addition
to the bare Coulomb potential, has to be included as well,
\begin{equation}\label{eq4}
V(r)=V_{n}(r)+V_{e}(r)=\frac{Ze}{r}+V_{e}(r)
\end{equation}

For low energy reactions the electronic part of the potential,
$V_{e}(r)$, has a small but important contribution to the total
potential. This contribution is amplified by the exponential
factor of the transparency.

In general, the electronic potential generated in a metallic
lattice by the conduction electrons is a slowly varying function.
Even for low energy particles the CTP of the projectile is so
close to the target nucleus that one can fairly assume that along
the integration path in (\ref{eq1}) the electronic potential
practically equals its value at the target nucleus, $V_{e}(r)\cong
V_{e}(0)$. When this approximation is inserted back into
(\ref{eq1}) it gets the form,
\begin{equation}\label{eq5}
T(E_{k})\cong \exp \left\{ -2\int_{r_{N}}^{b}\sqrt{\frac{2M}{\hbar
^{2}}\left[ \frac{Zz\,e^{2}}{r}+zeV_{e}(0)-E_{k}\right]
\,}dr\right\}
\end{equation}

This is equivalent to a reaction in a pure Coulomb potential with
the kinetic energy, $E_{k}$, replaced by $E_{k}+U_{e}$, where
\begin{equation}\label{eq6}
U_{e}=ze\,|V_{e}(0)| ~~~~~~~~(V_{e}(0)<0)
\end{equation}

The authors of Ref. \cite{rai04} tried to explain the experimental
results by means of the Debye statistical model, which neglects
the electrons degeneracy. A difficulty of their treatment is also
the point that under the experimental conditions the Debye radius
is significantly smaller than the atomic radius, therefore the
basic assumptions of the model are not satisfied.

The aim of the present paper is to compute the electrons spatial
distribution in insulator and metallic environments for the
$d+d\rightarrow t+p$ reaction $(Z=z=1)$, using a three-dimensional
Thomas-Fermi (TF) model, which accounts for the electrons'
degeneracy, therefore presumably better describing the experimental
conditions in a solid host of local high
density conditions ("strongly-coupled plasma").
Using the TF model we try
to infer the change in the transparency between metallic and
insulating environments. To illustrate our method, we shall focus
on the case of deuterium embedded in a copper lattice for which
experimental results are available \cite{rai04}.

\section{The model}\label{sec:2}
\subsection{Basic data}\label{subsec:2.1}
In the experiment, Ref. \cite{rai04}, the deuterium atoms consist
only a small part of the target - about 11 copper atoms for each
deuterium atom \cite{rai04} - one can, therefore, safely assume
that the presence of the deuterium atoms does not significantly
modify the copper lattice properties.

Our first step is to find the volume available for the deuterium
atoms. This is carried out by means of the QEOS method
\cite{mor88}, which is frequently used to find the
equation-of-state of various materials, and was found to give
accurate results. For our purposes its main advantage is that QEOS
can provide the volume per atom separately for each component in a
mixture of materials. Assuming a deuterium/copper solid material
with 9\% deuterium and 91\% copper at 8.93 $g/cm^{3}$ (solid
copper specific gravity), QEOS predicts that the volumes of the
deuterium (copper) atoms in the target are $
Vol_{D\,}\,(Vol_{Cu})=1.79\,(11.7)^{.}10^{-24}\;\,cm^{3}/atom$,
\textit{i.e.} , the average volume available for a copper atom is
by a factor of $\sim $ 6.5 larger than that of a deuterium atom.
If these atomic volumes are assumed to have the shape of a
spherical enclosure, called in the following the \textit{Ion
Sphere }(IS)\textit{, }then the corresponding IS radii are $
R_{i,D}\,(R_{i,Cu})=0.717\,(1.36)\,10^{-8}\,cm.$

The atomic structure of copper is $[Ar]\,3d^{10}\,4s$. In a pure
copper metal lattice there are at most 11 electrons per atom in
the conduction band. Their average density is
$n_{e}=11\,/\,11.7^{.}10^{-24}=9.4^{.}10^{23}
\,electrons\,/\,cm^{3}$. If the small amount of deuterium atoms
does not significantly change this distribution, then the same
density prevails within the deuterium IS as well, generating
$n_{e}Vol_{D\,}=9.4^{.}10^{23} \times 1.79^{.}10^{-24}=1.7$ extra
electrons inside the deuterium IS. Thus, together with its own
electron, the deuterium IS contains, on the average, $ N_{e}=2.7$
electrons. Changing the Cu/D target density by $\pm 10\%$ changes
this quantity only by $\pm $5\%. In our computations we have used
$N_{e}$, the number of electrons in the deuterium IS, as an
adjustable parameter.

\subsection{The modelling of the electronic potential in insulating and
metallic environments}\label{subsec:2.2} The electronic potential
of a free deuterium atom in high density insulating environment
was calculated by the computer program RELDIR, which solves the
relativistic Dirac equation of a deuterium (or any other) atom in
a finite radius Ion Sphere with any number of free electrons
within the IS. The results of this computation are the bound and
free electrons wavefunctions, their spatial density, and the
corresponding potentials.

As a first step, the computation was carried out for a solid
density deuterium lattice, which is known to be an insulator,
(0.202$\,g/cm^{3},$ 6.08$^{.}10^{22}\,atoms\,/\,cm^{3}$,
$R_{i}=1.58^{.}10^{-8}cm$). From the result of RELDIR for the
electronic potential one obtains $U_{e}=20\,eV$, in accordance
with the experimental results of Ref. \cite{rai04} in an
insulating environment. This result was used as the basic
reference for comparison.

The evaluation of the electronic potential near a deuterium atom
embedded in a metallic lattice was carried out by means of the
Thomas-Fermi (TF) model in conjunction with the Born-Oppenheimer
(BO) approximation. Owing to the high density environment, a
Fermi-Dirac statistics for the electrons, as used by the TF model,
provides a better approximation for the electrons spatial and
energy distributions inside the copper lattice than the Debye
model \cite{rai04,rai02} which is more appropriate for
weakly-coupled low-density matter. In this context, the physical
meaning of the BO approximation is that nearby electrons rapidly
adjust their local distribution to any change in the nuclei
positions, so that the electron distribution depends only on the
instantaneous distance between the two nuclei, but not on their
relative motion.

The TF model for the electrons consists of a set of three
equations. The first one is the Poisson equation,
\begin{equation}\label{eq7}
\overrightarrow{\mathbf{\nabla }}V_{e}(\mathbf{r})=4\pi
e\,n_{e}(\varepsilon_{F};\mathbf{r})
\end{equation}
whose solution is the electronic potential, $V_{e}(\mathbf{r})$,
when the electrons spatial density,
$n_{e}(\varepsilon_{F};\mathbf{r}),$ is known. In (\ref{eq7})
$\varepsilon_{F}$ is the Fermi energy. The second equation is the
Fermi-Dirac distribution of the electrons inside the IS,
\begin{eqnarray}\label{eq8}
n_{e}(\varepsilon_{F};\mathbf{r})=\frac{1}{2\pi ^{2}}\,\left(
\frac{2m_{e}\,kT}{\hbar ^{2}}\right) ^{3/2}\, \nonumber \\ \times
F_{1/2}\left( \frac{\varepsilon
_{F}+eV(\mathbf{r)}}{kT};\left\vert
\frac{eV(\mathbf{r)}}{kT}\right\vert \right)
\end{eqnarray}
which provides the electron density as function of the total
potential $V( \mathbf{r})=V_{n}(\mathbf{r})+V_{e}(\mathbf{r}).$ In
(\ref{eq8}), $kT$ \ is the temperature of the target material
($=300\,K)$ in energy units, and,
\begin{equation}\label{eq9}
F_{1/2}\left( x;\beta \right) =\int_{\beta }^{\infty }\frac{
y^{1/2}\,dy}{1+\exp \left\{ y-x\right\} }
\end{equation}
is the \textit{incomplete Fermi-Dirac integral } \cite{sal98}. The
boundary condition applied to Eqs. (\ref{eq7}) and (\ref{eq8}) is
Gauss' theorem, $\oint_{IS\,surface}\mathbf{E}
^{.}\mathbf{dS}=4\pi e\,N_{e}$. Finally, the Fermi energy,
$\varepsilon_{F}$, is computed from the condition that the total
charge inside the IS equals $ N_{e}$,
\begin{equation}\label{eq10}
N_{e}=\int_{IS\;\mbox{volume}}n_{e}(\varepsilon_{F};\mathbf{r}
)\,d^{3}r
\end{equation}

Simultaneous solution of Eqs. (\ref{eq8})-(\ref{eq10}) yields the
electronic potential, $ V_{e}(\mathbf{r})$, the electron density,
$n_{e}(\varepsilon_{F};\mathbf{r} ) $, and the Fermi energy,
$\varepsilon_{F}$. When these solutions are inserted back into
(\ref{eq1}) and (\ref{eq4}), one finds the CTP, $b(N_{e},E_{k})$,
and the transparency, $T(N_{e},E_{k})$ as function of the number
of electrons inside the IS and the incident particle's kinetic
energy.

\subsection{Computational details}\label{subsec:2.3}
The solution of (\ref{eq7}) can be rewritten as \cite{sal94},
\begin{eqnarray}\label{eq11}
V_{e}(r,\theta ,\varphi ) = -e\int \int
\int_{IS\;\mbox{volume}}\frac{ n_{e}(r^{\prime },\theta ^{\prime
})}{\left\vert \mathbf{r-r}^{\prime }\right\vert }d^{3}r \\
    = -e\int_{r_{N}}^{R_{i}}r^{\prime 2}\,dr^{\prime
}\,\int_{\theta =0}^{\pi }\sin \theta ^{\prime }\,d\theta ^{\prime
}\, \nonumber \\ \times \int_{\varphi =0}^{2\pi }d\varphi ^{\prime
}\,\frac{n_{e}(r^{\prime },\theta ^{\prime })}{ \left\vert
\mathbf{r-r}^{\prime }\right\vert } \nonumber
\end{eqnarray}
Obviously, the electron distribution has cylindrical symmetry
around the line connecting the two nuclei, therefore
$n_{e}(r^{\prime },\theta ^{\prime })$ is independent of $\varphi
^{\prime }.$ Moreover, the charge distribution has also a
reflection symmetry around the plane perpendicular to this line at
halfway between the nuclei. Using these symmetry properties, the
integration over $\varphi ^{\prime }$ can be carried out
analytically, thereby reducing the triple integral in (\ref{eq11})
to a double one, see details in Ref. \cite{sal94}. Finally,
transforming the coordinate system center onto the target nucleus,
Eq. (\ref{eq11}) gets the form,
\small{
\begin{eqnarray}\label{eq12}
V_{e}(R,\mu ) = -4e\int_{r_{N}}^{R_{i}}R^{\prime 2}\, dR^{\prime
}\, \int_{\mu =  \mu _{\min }}^{1}d\mu ^{\prime }\,n_{e}(R^{\prime
},\mu ^{\prime })\, \nonumber \\
\times \left[\frac{1}{\sqrt{A_{-}+B}}K\left( \sqrt{\frac{
2B}{A_{-}+B}}\right) + \frac{1}{\sqrt{A_{+}+B}}K\left(
\sqrt{\frac{2B}{A_{+}+B }}\right) \right] \nonumber\\
\end{eqnarray}
}
\begin{eqnarray}\label{eq13}
 A_{\pm } = \left[ \left( b/2+R\cos \theta \right) \pm
\left( b/2+R^{\prime }\cos \theta ^{\prime }\right) \right] ^{2}
\nonumber \\
 + ~(R\sin \theta )^{2}+\left( R^{\prime
}\sin \theta ^{\prime }\right) ^{2}
\end{eqnarray}
\begin{equation}\label{eq14}
B=2R\,R^{\prime }\,\sin \theta \,\sin \theta ^{\prime }
\end{equation}

In (\ref{eq12}) $R$ is the radial distance of the field point from
the target nucleus, $\mu =\cos \theta $, $K(x)$ is the Jacobi
elliptic integral \cite{abr72}, and $\mu _{\min }=\max \left(
-b\,/\,2R^{\prime },\,-1\right) .$ The reduction of the triple
integral in (\ref{eq11}) to the shape of (\ref{eq12}) reduced the
computational resources and improved the accuracy of the numerical
procedure.

The numerical process was greatly complicated by the fact that the
CTP, $b$, not only depends on the electron density,
$n_{e}(\varepsilon_{F};\mathbf{r} ) $, but also determines it. It
was, therefore, necessary to carry out a doubly iterative method
for the computations. The internal iterations calculated the
electron density and potential through equations
(\ref{eq12})-(\ref{eq14}) and (\ref{eq8})-(\ref{eq10}). When these
have been found, a second iterative procedure was applied to solve
$b$ from the condition,
\begin{equation}\label{eq15}
E_{p}(b)=E_{k}
\end{equation}

This double iterative method was continued until convergence was
achieved for all the parameters.

\section{Results}\label{sec:3}
\subsection{The electron density}\label{sec:3.1}
\begin{figure}[h]
\centering
\includegraphics[width=\columnwidth]{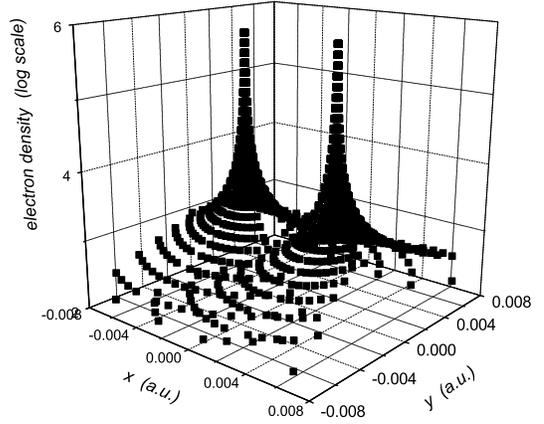}
\caption{The distribution of the electrons around the two nuclei.}
\label{fig:1}
\end{figure}

Figure~\ref{fig:1} shows the density of the electrons around the
incident and the target nuclei, when the center-of-mass energy of
the two nuclei is 4000~eV. The figure clearly shows the
polarization of the electrons near the nuclei. We recall that the
TF model predicts that the electron density diverges as $r^{-3/2}$
near the nucleus \cite{sal98}.

The TF model predicts an accumulation of the electrons along the
line connecting the nuclei, with a saddle point at halfway between
them, a fact which enhances the screening effect.

\subsection{The electronic potential}\label{sec:3.2}
\begin{figure}[h]
\centering
\includegraphics[width=\columnwidth] {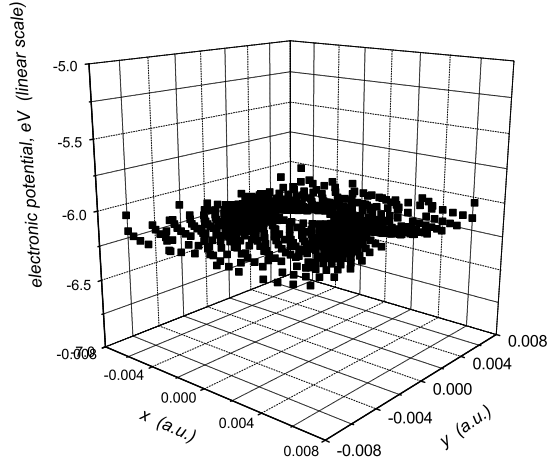}
\caption{The distribution of the electronic potential, $V_{e}(r)$,
around the two nuclei.} \label{fig:2}
\end{figure}

Figure~\ref{fig:2} shows the spatial distribution of the
electronic potential in the same region as in Fig.~\ref{fig:1}. In
the interesting part of the field, \textit{i.e.} in the space
around the two nuclei, the potential has very slow variation,
which justifies the approximation $V_{e}(r)\cong V_{e}(0)$, to an
accuracy of a few percents.

\begin{figure}[h]
\centering
 \includegraphics[width=3in] {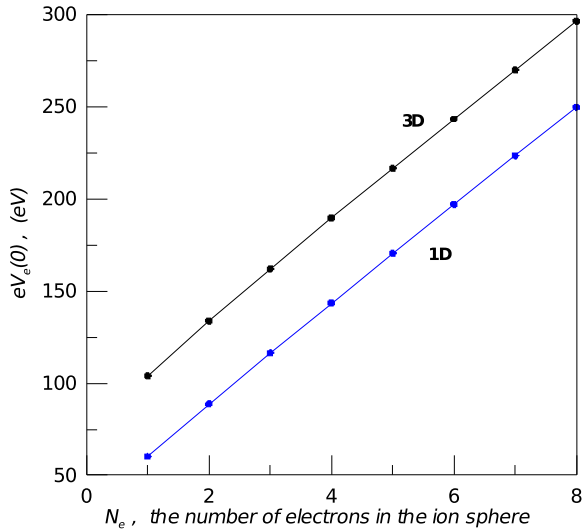}
\caption{The value of the electronic potential, $V_{e}(r)$, on the
target nucleus.}\label{fig:3}
\end{figure}

The value of the potential on the nuclei, $U_{e}=|eV_{e}(0)|$, is
displayed on Fig.~\ref{fig:3}. This is, in fact, the quantity that
is measured experimentally. The figure shows that $U_{e}$ fits
excellently a linear behavior as function of $N_{e}$, with a small
contribution from the incident energy,
\begin{equation}\label{eq16}
3D:U_{e} = |eV_{e}(0)| = 27.2\,N_{e}+80.3 +
3.1^{.}10^{-4}E_{k}\,~~eV
\end{equation}

The last term is always less than 4\% in the range of interest. In
order to compare the importance of a 3D modelling, we have
developed also a one-dimensional TF program that solves the same
problem in a radially symmetrical environment. The results of this
program is also plotted on Fig.~\ref{fig:3}. The TF 1D code
yielded,
\begin{equation}\label{eq17}
1D:U_{e}=|eV_{e}(0)|=27.2\,N_{e}\,+33.6 ~eV
\end{equation}

The difference between these two results originates from the fact
that the electrons accumulation between the nuclei in a 3D model
is larger than in the 1D case. This difference clearly indicates
the importance of a three-dimensional treatment of the problem.

For $N_{e}=3$, $E_{k}=4000\,eV$ \ Eq. (16) gives $U_{e}=163\,eV$.
This has to be compared to the experimental result, $U_{e}=470\pm
50\,eV$, which is by a factor of $2.9\pm 0.3$ larger than the
result of the TF 3D model. It has also to be compared to
$U_{e}=20\,\,eV$ as computed from RELDIR for the insulator case,
see above.

\subsection{The Fermi energy}\label{sec:3.3}
\begin{figure}[b]
\centering
 \includegraphics[width=3in] {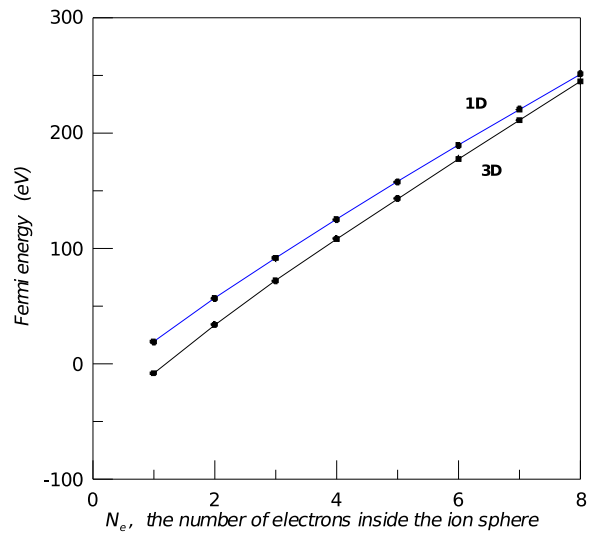}
\caption{Fermi energy as function of the number of electrons
inside the ion sphere, for 3D and 1D Thomas-Fermi
simulations.}\label{fig:4}
\end{figure}

The Fermi energy, $\varepsilon_{F}$, strongly depends on the
number of electrons in the IS, $N_{e}$, and is, of course,
independent on the incoming particle's energy. The behavior of
$\varepsilon_{F}$ vs. $N_{e}$ is illustrated in Fig.~\ref{fig:4}.
Within the range of our computations $\varepsilon_{F}$ turned out
to be a linear function of $N_{e}$,
\begin{equation}\label{eq18}
3D:\quad \varepsilon_{F}=-43.7+38.8\,N_{e}\qquad (eV)
\end{equation}
with a hint for a slight quadratic curvature. For the 1D case we
have found a significantly higher result,
\begin{equation}\label{eq19}
1D:\quad \varepsilon_{F}=-15.9+35.3\,N_{e}\qquad (eV)
\end{equation}

It should be emphasized, that the Fermi energy is positive (except
for the case of $N_{e}=1,$ namely, the case of an isolated
deuterium atom), and is much higher than the target temperature
($\varepsilon_{F}\gg kT=300\,K=0.025\,eV)$, for all the cases.
This means that at the temperatures of the experiments (room
temperature and below) there is no reason for measurable
temperature variations. In fact, we have repeated our computations
with $T=3000\,K$ , but this order of magnitude change in the
temperature modified the results by less than 0.5\% - as expected.

\subsection{The classical turning point}\label{sec:3.4}
The CTP fits, with excellent accuracy, a function of the form,
\begin{equation}\label{eq20}
b\,/\,a_{0}=\frac{e^{2}\,/\,a_{0}}{E_{k}+U_{e}}
\end{equation}
where $e^{2}/\,a_{0}=27.211\,eV$, ($a_{0}$ is the Bohr radius).
This has exactly the form of \ the Coulomb CTP, with the
projectile's kinetic energy modified according to Eq. (5). Fitting
the computational results to the form of (20) provides,
\begin{equation}\label{eq21}
U_{e}=27.7\,N_{e}+76.4+5.76^{.}10^{-4}\,E_{k}\quad eV
\end{equation}

The agreement of this $U_{e}$ with the values obtained from the
electronic potential, Eqs. (\ref{eq16}), can be regarded as very
good. The difference between the two results stems from the fact
that $|eV_{e}(\mathbf{r})|$ is not exactly constant along the line
connecting the two nuclei. And again, this result is in
discrepancy with the experimental value, but is much higher than
the insulator case.

\subsection{The transparency}\label{sec:3.5}
\begin{figure}[b]
\centering
 \includegraphics[width=3in] {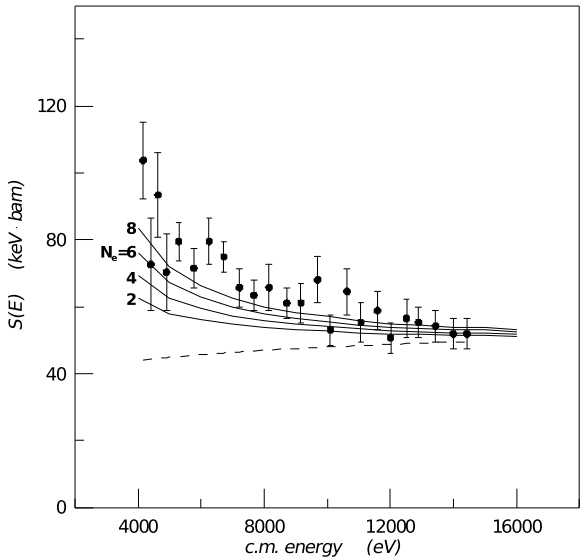}
\caption{Comparison between the results of the 3D TF model
for the astrophysical factor, S(E), of the reaction $d+d \rightarrow
t+p$, when the target nuclei are embedded in a copper substrate,
and the experimental results. The experimental points are taken from
\cite{rai04}. The dotted line presented the theoretically extrapolated
"bare nucleon" S(E) factor}.\label{fig:5}
\end{figure}

At low impact energies the screened astrophysical factor, $S(E)$,
is enhanced relative to the unscreened one by
 $T(E)/T_{C}(E)$ \cite{rai04}, see Eqs. (\ref{eq1}) and
(\ref{eq2}). Fig.~\ref{fig:5} shows our results for this factor as
function of the center of mass energy for the $d+d\rightarrow t+p$
reaction, when the target deuterium is embedded in a copper
lattice. The experimental results of Ref. \cite{rai04} are also
displayed in Fig.~\ref{fig:5}. The dashed curve represents the
bare $S(E)$ factor \cite{rai04}. While we note that the value
$N_{e}$ = 3 is a reasonable estimate for the conducting electrons in
the deuterium IS embedded in Cu, even for $N_{e}$ = 8 the 3-dimensional
TF model underestimates the experimental low-energy
increase of the astrophysical factor.
The low $U_{e}$ predicted by the model relative to the experiment is another
manifestation of the same fact.
In fact, one needs 11 conduction electrons inside the deuterium IS
to get agreement with the experimental results.

In the range of the energies used in our computations $U_{e}\leq
0.1E$. Denoting $\xi =U_{e}\,/\,E,$ $\xi $ can be assumed to be a
small quantity. Using first order expansion, an analytical
estimate can be developed for the change in the value of
$T(E)\,/\,T_{C}(E)$ caused by a change $b\rightarrow b\,/\,(1+\xi
)$ in the CTP, see Eq. (\ref{eq20}). The ratio of the screened to
Coulomb transparencies becomes,

\begin{equation}\label{eq22}
T(E)\,/\,T_{C}= exp\left\{ \delta G\right\}
\end{equation}
where,
\begin{equation}\label{eq23}
\delta G=G-G_{C}=2e^{2}\sqrt{\frac{2M}{\hbar ^{2}}}\frac{\xi
}{\sqrt{ E}}\left( \frac{\pi }{4}-\sqrt{\frac{\xi }{2}}\right)
\end{equation}
For $E=4000\,eV,$ $N_{e}=3$ and $U_{e}=163$ $eV$ this formula
predicts,
\begin{equation}\label{eq24}
T(E)\,/\,T_{C}= exp\left\{ \delta G\right\} =1.45
\end{equation}
in contrast to the experimental result of $T(E)\,/\,T_{C}\approx
2.0\pm 0.2\,.$

\section{Discussion}\label{sec:4}
In this paper we present computational results from a
three-dimensional TF model about the modification of low-energy
cross-section for the $d+d\rightarrow t+p$ reaction, when the
target nucleus is embedded in a copper lattice. Our computational
result of $U_{e}=163\,eV$ is lower by a factor of $\sim $3 from
the experimental result of $U_{e}=470\pm 50\,eV,$ but still is
substantially higher than the cross-section in an insulator (solid
deuterium lattice), $U_{e}=20\,\,eV.$

\begin{figure}[b]
\centering
\includegraphics[width=3in] {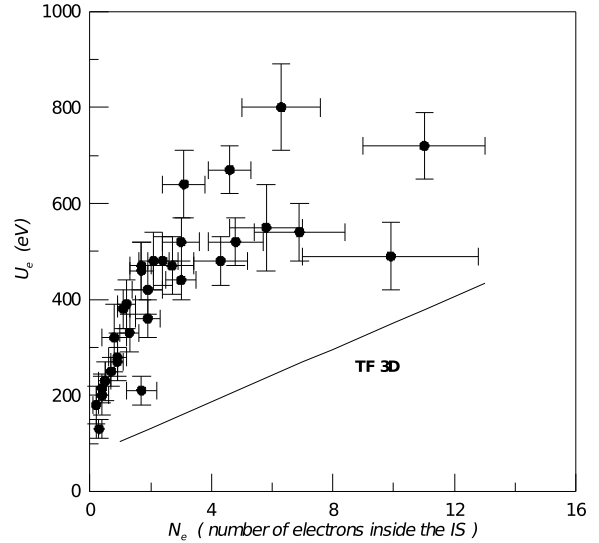}
\caption{The experimental values of $U_{e}$, drawn from the results of
 \cite{rai04}, compared to the predictions of the 3D TF model.}
 \label{fig:6}
\end{figure}

In order to see whether such a difference holds true for other
target lattices as well, in Fig.~\ref{fig:6} we plotted the
results of all the experiments in metal lattices published in Ref.
\cite{rai04} in comparison to ours. The 3D TF results in the
figure may be shifted $\pm 5\%$ up or down, due to differences in
the local IS volume of the deuterium in the various lattices.
Figure~\ref{fig:6} exhibits a consistent difference between the
computational and the experimental results. As the Thomas-Fermi is
a highly successful model, well fitted to high density matter,
this insistent difference is of some surprise.

Obviously, the central reason for the disagreement is the
electronic potential on the target nucleus, see Eq. (\ref{eq6}).
In our opinion, the semiclassical nature of the TF model cannot be
the reason for this difference, because the quantum mechanical
behavior of the conduction electrons seems to be unimportant in
the present problem.

On the other hand, the TF model assumption of a perfect spherical
symmetry, and the target nucleus location at the center of a
well-defined ion sphere, may oversimplify the real situation. In
the experiment, the deuterium nuclei are implanted into the metal
lattice by bombardment of $10\,keV$ deuterons into the metal foil
\cite{rai02}. It is well known that this technique does not
necessarily deposit the stopped deuterium into a spherically
symmetrical environment.

Another contribution to the discrepancy may be the local
conduction electron density fluctuations around the target
nucleus. As the transparency depends exponentially on the local
electron potential, a small number of nuclei, with an
instantaneous large electron screening, has greater influence on
the cross-section than the other nuclei with the average
electronic screening.

Finally, it is possible that QEOS underestimates the deuterium IS
volume under the specific experimental conditions. Larger IS
volume, with more conduction electrons, may have better agreement
with the experiment.

\end{document}